\documentclass[draft,wrr]{agutex}
 \usepackage{lineno}
 \linenumbers*[1]
\usepackage[pdftex]{graphicx}
\setkeys{Gin}{draft=false}
\authorrunninghead{L\'EGER, SAINTENOY and COQUET}
\titlerunninghead{Inverting Saturated Hydraulic Conductivity}
\authoraddr{Corresponding author: E. L\'eger,
Department of earth science, Universit\'e Paris Sud, Building 504, Orsay, 91405, France.
(emmanuel.leger@u-psud.fr)}
\begin{document}
\title{Saturated hydraulic conductivity determined by on ground mono-offset Ground-Penetrating Radar inside a single ring infiltrometer}
\authors{Emmanuel L\'eger,\altaffilmark{1}
 Albane Saintenoy,\altaffilmark{1} Yves Coquet,\altaffilmark{3}}

\altaffiltext{1}{Universit\'e Paris Sud, UMR 8148 IDES, Orsay, France.}
\altaffiltext{2}{Universit\'e Orl\'eans, ISTO/OSUC., Orl\'eans, France.}
\begin{abstract}
In this study we show how to use GPR data acquired along the
infiltration of water inside a single ring infiltrometer to inverse
the saturated hydraulic conductivity. We used Hydrus-1D to simulate
the water infiltration. We generated water content profiles at each
time step of infiltration, based on a particular value of the
saturated hydraulic conductivity, knowing the other van Genuchten
parameters. Water content profiles were converted to dielectric
permittivity profiles using the Complex Refractive Index Method
relation. We then used the GprMax suite of programs to generate
radargrams and to follow the wetting front using arrival time of
electromagnetic waves recorded by a Ground-Penetrating Radar (GPR).
Theoretically, the 1D time convolution between reflectivity and GPR
signal at any infiltration time step is related to the peak of the
reflected amplitude recorded in the corresponding trace in the
radargram. We used this relationship to invert the saturated hydraulic
conductivity for constant and falling head infiltrations. We present
our method on synthetic examples and on two experiments carried out on
sand soil. We further discuss on the uncertainties on the retrieved
saturated hydraulic conductivity computed by our algorithm from the
van Genuchten parameters.
\end{abstract}

\begin{article}

\section{Introduction}

Soil hydraulic properties, represented by the soil water retention
$\theta(h)$ and hydraulic conductivity $K(h)$ functions, dictate water
flow in the vadose zone, as well as partitioning between infiltration
and runoff. Their evaluation has important implications for modeling
available water resources and for flood forecasting. It is also
crucial in evaluating the dynamics of chemical pollutants in soil
and in assessing the potential of groundwater pollution.\\
Soil hydraulic functions can be described by several mathematical expression~\citep{Kosugi2002}, among them the van Genuchten function~\citep{vanGenuchten1980}. 
 The determination of the parameters defining the van Genuchten soil water
retention function~\citep{vanGenuchten1980} is usually done using
laboratory experiments, such as the water hanging
column~\citep{Dane2002}. 

The hydraulic conductivity function 
can be estimated either in the laboratory, or in situ using
infiltration tests. Among the large number of existing infiltration
tests~\citep{Angulo2002}, the single~\citep{Muntz1905} or double ring
infiltrometers~\citep{Boivin1987} provide the field saturated hydraulic
conductivity by applying a positive water pressure on the soil surface,
while the disk infiltrometer~\citep{Perroux1988,Clothier1981} allows to
reconstruct the hydraulic conductivity curve, by applying different water
pressures smaller than or equal to zero. For infiltration tests, the
volume of infiltrated water versus time is fitted to infer the soil
hydraulic conductivity at or close to saturation. These tests are
time-consuming and difficult to apply to landscape-scale forecasting
of infiltration. Furthermore, their analysis involve various
simplifying assumptions, partly due to the ignorance of the shape of
the infiltration bulb. This lack of knowledge on the form of the infiltration bulb has to be
filled to get accurate informations on the soil water retention
$\theta(h)$ function and consequently on hydraulic conductivity $K(h)$
function. This can be done by water content sensing. 

Vereecken~\citep{Vereecken2008} and Evett and Parkin~\citep{Evett2005} give a state of the art on the different techniques
available for soil moisture measurements. Among the large
panel presented, geophysical methods take an important part, mainly
because they are contact free and/or easy to use. The most commonly used
hydro-geophysical methods are electrical resistivity
measurements~\citep{Goyal2006,Zhou2001} and electromagnetic
methods~\citep{Sheets1995,Akbar2005}. This paper focuses on
the use of Ground-Penetrating Radar (GPR) as a tool for monitoring
water infiltration in soil.

For few decades GPR has been known as an accurate method to highlight water
variation in soils~\citep{Huisman2003,Annan2005}. Different
techniques are available in the literature for monitoring water content in soils using GPR. Tomography imaging between
boreholes during an infiltration has been done by Binley~\citep{Binley2001} and Kowalsky~\citep{Kowalsky2005} among others. Many advances were done during
the last years on Off-Ground GPR using full waveform inversion, for
instance to invert soil hydraulic properties (Lambot~\citep{Lambot2006,Lambot2009} and
Jadoon~\citep{Jadoon2012}). Grote~\citep{Grote2002} and Lunt~\citep{Lunt2005} used two-way travel time variations from a reflector
at a known depth to monitor water content variation with time. Finally,
multi-offset GPR survey techniques, i.e. CMP\footnote{Common MidPoints} or WARR\footnote{Wide-Angle Reflection- Refraction}, were carried out
during infiltration processes in the works of Greaves~\citep{Greaves1996} or Mangel~\citep{Mangel2012}.

The work presented here is based on mono-offset monitoring of infiltration with
on-ground surface GPR as related by Haarder~\citep{Haarder2011}, Moysey~\citep{Moysey2010}, Lai~\citep{Lai2012}, Dagenbach~\citep{Dagenbach2013}
and Saintenoy~\citep{Saintenoy2008}. Haarder~\citep{Haarder2011} used a constant offset on-ground GPR coupled
with dye tracing to exhibit preferential flows. They found that a GPR
was able to map deep infiltration comparing to dye tracer, but
they did not manage to resolve the infiltration patterns
(by-pass flow, fingering...). Moysey~\citep{Moysey2010} studied the infiltration inside a sand box from
the surface with on-ground GPR. He used the reflection from the
wetting front as well as from the ground wave and the bottom of the box, to
monitor the water content. He also modelled his experiment and estimated the van Genuchten parameters using semblance analysis. As
L\'eger~\citep{Leger2013}, he found that the most
poorly constrained parameter was $n$. Lai 
~\citep{Lai2012} used a joint time frequency analysis coupled with
grayscale imaging to measure infiltration and drainage in controlled
conditions in laboratory. They were able to follow the peak frequency
of the GPR wavelet associated with the wetting front using time
frequency analysis and then determined the rate of water infiltration
in unsaturated zone. Saintenoy~\citep{Saintenoy2008}
monitored the wetting bulb during an infiltration from a Porchet
infiltrometer. They were able to identify the dimension of the bulb
with time and good agreement was found with modelling.

On the continuity of those studies, we present a method for monitoring
the wetting front during infiltration using on-ground GPR with
fixed offset inside a ring infiltrometer. The objectives of this paper
were i) to check if the proposed method is accurate enough to monitor wetting
front during infiltration with different boundary conditions, ii)
to invert saturated hydraulic conductivity using the model of
Mualem-van Genuchten~\citep{Mualem1976,vanGenuchten1980} , and iii) to analyze the uncertainties using a simplified MC uncertainty analysis.
The method has been tested on synthetic examples and on two field data
sets.

\section{Background}

\subsection{Unsaturated Flow Equation}

In this study we consider one-dimensionnal vertical water flow in a soil, described by the one-dimensional Richard's
equation~\citep{Richards1931}. Its expression in term of water
content is
\begin{eqnarray}
\frac{\partial\theta}{\partial t}=\frac{\partial K(\theta)}{\partial
  z} +\frac{\partial}{\partial z}\left[D(\theta)\frac{\partial\theta}{\partial z}\right]~,
  \label{Rich}
\end{eqnarray}
where $K(\theta)$ is the hydraulic conductivity as a function of water
content, and $D(\theta)$ is water diffusivity (Childs
and Georges-Collis~\citep{Childs1950}), expressed in terms of water content as $D(\theta)=K(\theta)\frac{\partial h}{\partial\theta}$.

\subsection{Hydraulic Properties Functions}

Several mathematical functions exist to model the hydraulic
properties of porous media~\citep{Kosugi2002}.
We chose the van Genuchten model~\citep{vanGenuchten1980}
with the relation of Mualem~\citep{Mualem1976}, giving the following
expression for the water retention
curve:
\begin{eqnarray}
\theta(h)=\theta_r+(\theta_s-\theta_r)(1+(\alpha h)^n)^{\frac{1}{n}-1}~,
\end{eqnarray}
where $\theta_s$ is the saturated water content, $\theta_r$, the
residual water content, and $\alpha$ and $n$, two fitting parameters
which are respectively linked to the matric potential and the slope of
the water retention curve at the inflexion point. The hydraulic conductivity function is 
described by
\begin{eqnarray}
K(\theta)=K_s\Theta^{\lambda}\left[1-\left[1-\Theta^{\frac{n}{n-1}}\right]^{\frac{n}{n-1}}\right]^2~,
\end{eqnarray}
with $K_s$ the saturated hydraulic conductivity,
$\Theta=\frac{\theta-\theta_r}{\theta_s-\theta_r}$ the effective
saturation and $\lambda$ a factor that accounts for pore
tortuosity. The $\lambda$ parameter has an
influence on the shape of the hydraulic conductivity function. However
in this study we concentrated on the inversion of only one parameter, the
saturated hydraulic conductivity. We fixed $\lambda$ equal to 0.5 as
reported in~\citep{Mualem1976}.

\subsection{Petrophysical Relationships}

Several empirical and conceptual relationships exist to convert soil dielectric
permittivity to volumetric water content.
Using the fact that the experiments presented here have been made in a quarry of Fontainebleau sand, considered as pure
silica, we used the CRIM relation~\citep{Birchak1974,Roth1990}, which
relates the relative dielectric permittivity of bulk media,
$\varepsilon_b$, to the volumetric summation of each components of it.
Thus for a tri-phasic medium comprising water, air and silicium, we obtain
\begin{eqnarray}
\sqrt{\varepsilon_b}=\theta\sqrt{\varepsilon_w}+(1-\phi)\sqrt{\varepsilon_s}+(\phi-\theta)~,
\label{CRIM}
\end{eqnarray}
where $\varepsilon_w=80.1,\ \varepsilon_s=2.5$ are respectively the relative
dielectric permittivity of water and silica, $\phi$ the porosity and
$\theta$ the volumetric water content.

\subsection{Dielectric Permittivity Versus Electromagnetic Wave Velocity}

Surface GPR consists in a transmitting antenna, being a dipole,
positioned on the surface, that emits short pulses of spherical
electromagnetic (EM) wave in response to an excitation current source,
and a receiving antenna, also located at the surface, which converts the
incoming EM fields to an electrical signal source to be treated.
Following the works of Annan~\citep{Annan1999},
the velocity of electromagnetic waves is
\begin{eqnarray}
v=\frac{c}{\sqrt{\varepsilon^{\prime}  \mu_r\frac{1+\sqrt{1+tan^2\delta}}{2}}}~,
\label{V}
\end{eqnarray}
where $\delta$ is the loss factor as a function of the dielectric
permittivity, frequency and electrical conductivity, $\varepsilon'$ is
the real part of the relative dielectric permittivity, $\mu_r$ the
relative magnetic permeability and $c$ is the velocity of EM waves in
air equal to $0.3\ m/ns$. Considering the case of non magnetic soil
with low conductivity, in the range of 10~MHz to 1~GHz, the real part
dominates the imaginary part of the dielectric permittivity and
neglecting Debye~\citep{Debye1929} effect, equation (\ref{V}) reduces
to:
\begin{eqnarray}
v=\frac{c}{\sqrt{\varepsilon^{\prime}}}~.
\label{Huism}
\end{eqnarray} 
We used this equation to compute the travelling time of an EM wave
through a layer of soil of known thickness with a given dielectric
permittivity.

\subsection{Electromagnetic Modelling}
Numerous techniques are available for simulating GPR data, e.g. ray-based methods (e.g. Cai and McMechan~\citep{Cai1995} or Sethian and Popovici~\citep{Sethian1999}), time-domain finite-difference full-waveform methods (e.g. Kunz and Luebbers~\citep{Kunz1996} or Kowalsky~\citep{Kowalsky2001}), or finite differences time domain (FDTD) (e.g. Irving and Knight~\citep{Irving2006}).  We used the GprMax 2D codes of Giannopoulos~\citep{Giannopoulos2005}, which uses FDTD modelling to solve the maxwell equations in 2 dimensions.

\section{Materials and Methods}

\subsection{Experimental Set-up}

\begin{figure}[htpb]
\noindent\includegraphics[width=7cm]{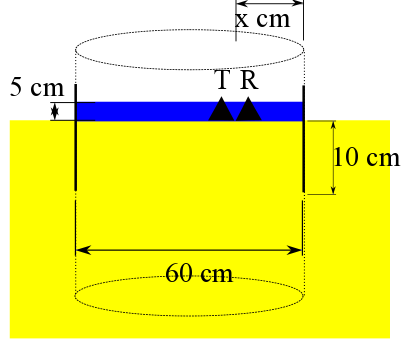}
\caption{Experimental set up at its initial state.}
 \label{Schema}
 \end{figure}

We studied infiltration of a 5-cm thick water layer inside of a single
ring infiltrometer in a sandy soil. The scheme of the apparatus is
presented in Figure \ref{Schema}. The single ring infiltrometer was a
1-mm thick aluminum cylinder with a 60-cm diameter, approximately 20-cm
high, buried in the soil to a depth of 10 cm. GPR antennae
(namely the transmitter T and the receiver R) were set up at a variable distance
from the edge of the cylinder, noted $X,$ in Figure \ref{Schema}. In
all our field experiments, we used a Mala RAMAC system with antennae
centered on 1600 MHz, shielded at the top. The inner part of the cylinder was covered with a plastic waterproof sheet. This allowed us to fill the cylinder with water and create an initial
5-cm thick water layer, while preventing infiltration into the sand
before starting data acquisition. The beginning of the acquisition was
launched by pulling away the plastic sheet to trigger water
infiltration. The GPR system was set to acquire a trace every 10 s.
With this apparatus, we performed two types of infiltration: i) a falling head infiltration consisting of pulling away the plastic sheet
and leaving water to infiltrate into the sand freely with no
additional refill, and ii) a constant head infiltration, when water
was continuously added to the ring to maintain a 5-cm thick water
layer during the infiltration experiment. In the following examples, we
will show that GPR data acquired every 10 s during the
infiltration experiment can be used to estimate the saturated soil hydraulic
conductivity, $K_s$. In all GPR data presented below, we subtracted the average trace and applied an Automatic Gain Control (AGC) to the data in order to make them clearer.
The van Genuchten parameters, $\alpha,\ n,\ \theta_r,\ \theta_i$ of the sand have been determined in laboratory by several
classical hanging water column experiments. We assumed arbitrarily a $5\ \%$ uncertainty for all the measured
parameters. The sand was considered homogeneous. Its initial water content, $\theta_i$, and  porosity, $\phi$, of the soil were determined using gravimetric measurements on field samples.

\subsection{Modelling}
Infiltration experiments were simulated by solving Richards equation (Eq.~(\ref{Rich}) ) using Hydrus-1D. The soil profile was 50 cm deep, assumed to be homogeneous, and divided into 1001 layers. We used either an atmospheric boundary condition (BC) with no rain and no evaporation at the soil surface, for the falling head infiltration, or a constant pressure head of 5 cm to the top
node, for the constant head infiltration, and for both case free drainage BC at the bottom. To simulate the 5-cm layer of water, the initial condition was set to a 5 cm pressure head in the top node. We simulated the first 10 minutes of the experiment with a time step of 10 s, i.e., with 60 water content snapshots. Using the CRIM relation (Eq.~\ref{CRIM}), each water content snapshot was converted to permittivity profiles (made of 1001 points), considering a three-phase media: sand (considered as pure silica), water, and air. Each one of these permittivity profiles were the input for the GprMax2D program~\citep{Giannopoulos2005}. GprMax2D gave simulated GPR monitoring of the infiltration process. We then picked the maximum amplitude of the signal to get the Two Way Travel (TWT) time of the wetting front reflection.

\subsection{Inversion Algorithm}
\subsubsection{Convolution Algorithm}
Our inversion algorithm was based on the comparison between the arrival
times of the wetting front reflection observed in the radargrams acquired
during the water infiltration experiment
and the arrival times of these reflections computed from
the theoretical water content profiles modeled by Hydrus-1D. If a suitable
relationship between water content and dielectric permittivity is
known, water content profiles,
obtained by the resolution of the Richards~\citep{Richards1931} equation
(done by Hydrus1D in our case), can be transformed to a 2D series of reflection
coefficients:
\begin{eqnarray}
R_{i,t}=\frac{\sqrt{\varepsilon_{i+1,t}}-\sqrt{\varepsilon_{i,t}}}{\sqrt{\varepsilon_{i+1,t}}+\sqrt{\varepsilon_{i,t}}}~,
\label{refl}
\end{eqnarray}
where $\sqrt{\varepsilon_{i,t}}$ and $\sqrt{\varepsilon_{i+1,t}}$ are
the relative dielectric permittivity at the infiltration time $t$ for
two successive model cells centered at depth $z_i$ and $z_{i+1}$.The effective depth where the reflection coefficient is
calculated is $z_R=\frac{z_i+z_{i+1}}{2}$. Knowing the dielectric
permittivity of each layer of the profile, the
electromagnetic wave velocity (Eq.~\ref{Huism}) and travel time can be computed.
The travel time is used to interpolate reflection coefficients to a
constant sampling interval. We used this depth to time conversion to
compute a Ricker signal in this time interval. The center frequency of the Ricker was set to 1000 MHz, central frequency of the GPR signal recorded on the field. We derived it twice with respect to
time to simulate the transformation made by the emitter and the
receiver in real antennae. We then performed the convolution between
this pseudo-GPR signal and the reflectivity to obtain
\begin{eqnarray}
O(t)=R(t)\ast\frac{\partial^2}{\partial t^2}I(t),
\end{eqnarray}
where $O(t)$ is the output signal, $R(t)$ is the reflectivity and
$I(t)$ is the input source of the antenna.

Some remarks have to be made about the comparison between 1D-temporal convolution and real electromagnetic signal. First of all, our
inversion algorithm is based on the assumption that soil can be
represented as a stack of homogeneous layers. The assumption of horizontal
interfaces forces the reflection coefficient (equation
(\ref{refl})) to be expressed as a normal incidence case. Secondly, we considered that the 2-D plane waves computed by FDTD algorithm (modelling) and 3-D plane waves (experiments) could
be treated as a 1-D temporal convolution. Third we neglect relaxation effects occurring when
propagating an electromagnetic wave in water saturated sand.

\subsubsection{Inversion Procedure} 
We used the TWT time obtained from the radargram (modelled or experimental) as data to be fitted to derive the saturated hydraulic conductivity,
assuming the other 4 van Genuchten parameters and initial water content were known. Using Hydrus-1D, we generated 60 water content snapshots  using the saturated hydraulic
conductivity in the range from $0.01$ to $1\ cm/min$, with a step of $0.001\ cm/min$. For each value of $K_s$, we calculated the TWT time using our
convolution algorithm and we computed the Root Mean Square Error (RMSE) between these times and the data as an objective function, to be computed as function of saturated hydraulic conductivity. The $K_s$ value which corresponds to the minimum of the objective function was used as inverted value.

\section{Falling Head Infiltration Experiment}

\subsection{Numerical Example}

\subsubsection{Forward modelling}

\begin{figure}[htpb]

\noindent\includegraphics[width=15cm]{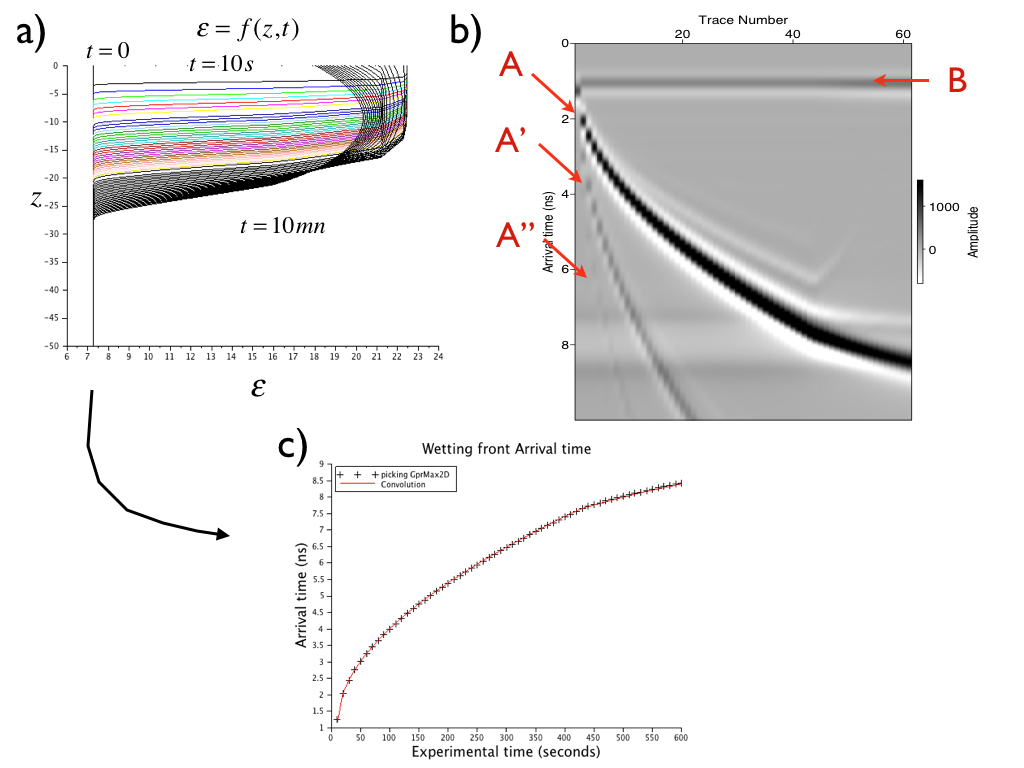}

\caption{Falling head infiltration from a 5-cm thick water layer. a)
  Permittivity profiles: each curve is plotted every 10 s.
  b) Radargram simulated with GprMax2D; reflection A is coming from
  the wetting front, B is the direct wave, A' and A'' are multiples of
  reflection A. c) TWT time computed by the convolution algorithm from the
  permittivity profiles (plain red line) and TWT time obtained by picking of A peak in fig b).}

\label{algo1}
 \end{figure}

The set of hydrodynamical parameters used for this numerical
example is presented in Table~\ref{param}. The permittivity profiles, resulting from water content conversions from Hydrus-1D to permittivity and which were used as input of GprMax2D program~\citep{Giannopoulos2005} are presented in Figure~\ref{algo1}-a. The simulated GPR monitoring of
the infiltration process is shown in Figure \ref{algo1}-b. The
horizontal axis is the number of traces simulated with GprMax2D, two
traces being separated by 10 seconds, as permittivity profiles are. The
vertical axis is the TWT  time of the EM wave coming back to the receiver.

On the profile presented in Figure~\ref{algo1}-b, we denote one
particular reflection, labeled A. Its arrival time is increasing as
the wetting front moves deeper. This reflection is interpreted as
coming from the wetting front. The reflections labeled $A^{\prime}$
and $A^{\prime\prime}$ are primary and secondary multiples of
reflection $A$. The reflection labeled $B$ is the wave traveling in
air directly between the two antennae. After the 40$^{th}$ trace, the
5-cm layer of water has been infiltrated, and drainage is starting. As
a consequence, the permittivity of the upper part of the medium
decreases and the velocity increases (Eq.~\ref{Huism}). The TWT time of reflection $A$ increases more slowly, creating a change
of slope in the reflection time curve (Fig. \ref{algo1}-b). In Figure
\ref{algo1}-c, we display two curves: the TWT time of the maximum peak of reflection $A$ (obtained from Figure \ref{algo1}-b) and the TWT time
calculated by the convolution Algorithm.

 The result of the convolution algorithm is in good agreement with the GprMax2D modelling.

\subsubsection{Inverse Modeling}

We used the TWT time obtained from the radargram of Figure \ref{algo1}-b
as data to be fitted to derive the saturated hydraulic conductivity,
assuming the other 4 van Genuchten parameters and initial water content were known (see Table \ref{param}).
 The RMSE was minimized for
$K_s=0.121\ cm/min$, which has to be compared with the value set as
input, i.e., $K_s=0.120\ cm/min$. This result confirms the ability of our algorithm to invert saturated hydraulic conductivity.

\subsection{Field experiment}

\subsubsection{Experimental Data and its Analysis}

\begin{figure}[htpb]

\noindent\includegraphics[width=15cm]{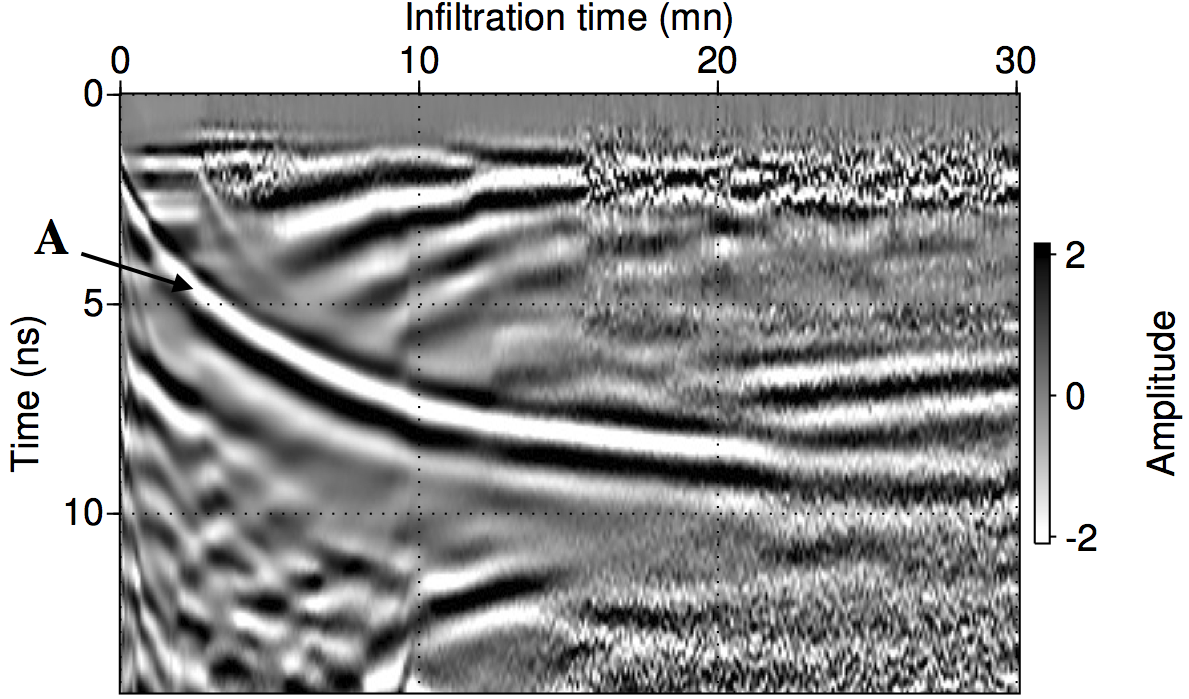}

\caption{Experimental GPR data acquired during the falling head
  infiltration (using a 5-cm initial water layer). Reflection A is the
  reflection coming from the wetting front}

\label{R1}
\end{figure}

The experiment took place in a quarry of Fontainebleau sand in
Cernay-La-Ville (Yvelines, France). The middle of the antennae was
positioned 11 cm away from the cylinder wall ($X = 11\ cm$ in
Fig.~\ref{Schema}). The 5-cm water layer was fully infiltrated after about
10 minutes, although in certain areas of the soil surface this time
has been slightly shorter. The sand parameters measured by the hanging water column are given in Table~\ref{param} and initial volumetric water content is $\theta_i=0.09\pm0.01\ cm^3/cm^3$. The recorded GPR data are shown in
Fig.~\ref{R1}. 
 In this profile, we denote three particular reflections. The one
interpreted as coming from the infiltration front, labeled A, is
visible during the first 30 minutes of the acquisition, with an
arrival time varying from 2 ns down to 9 ns. The other reflections come from the cylinder and are 
interpreted in~\citep{Leger2012}. We determined the arrival time of the
A reflection peak and inverted the saturated hydraulic conductivity
using the same algorithm as for the synthetic case. We obtained the
minimum of the objective function for $K_s=0.120\ cm/min$. In parallel,
we also carried out disk infiltrometer experiments, using the
multi-potential method~\citep{Ankeny1991,Reynolds1991}. We obtained a
value of the saturated hydraulic conductivity of
$K_{Disk}=0.108\pm0.01\ cm/min$.

\subsubsection{Uncertainty Analysis}
We attempted to evaluate the uncertainty in the saturated hydraulic
conductivity retrieved from GPR data fitting by using a modified Monte
Carlo method. We qualified this method as ``modified Monte Carlo'' in the
sense that it is nor the Tarantola method~\citep{Tarantola1987} and neither
the adaptive method proposed by the Guide to the expression of
uncertainty in measurement~\citep{GUM} published by the Joint Committee
for Guides in Metrology (JCGM). We consider five major uncertainty
sources, four from the van Genuchten parameters, $\alpha$, $n$, $\theta_r$,
$\theta_s$ and one from the initial water content $\theta_i$. We do assume
that all uncertainties can by described by gaussian distribution
probability function centered on the value found by several water hanging column experiments with a standard deviation of $5\ \%$ of
this value. With this definition we obtained the following set of a
priori density function for experimental case:
$\mathcal{N}_{\alpha}(\alpha^{\mu}=0.023\ cm^{-1},\ \alpha^{\sigma}=0.001\ cm^{-1})$,
$\mathcal{N}_{n}(n^{\mu}=6.7,\ n^{\sigma}=0.3)$,
$\mathcal{N}_{\theta_r}(\theta_r^{\mu}=0.062\ cm^3/cm^3,\ \theta_r^{\sigma}=0.001cm^3/cm^3)$,
$\mathcal{N}_{\theta_s}(\theta_s^{\mu}=0.39\ cm^3/cm^3,\ \theta_s^{\sigma}=0.01\ cm^3/cm^3)$, and $\mathcal{N}_{\theta_i}(\theta_i^{\mu}=0.09\ cm^3/cm^3,\ \theta_i^{\sigma}=0.01\ cm^3/cm^3)$, 
where the $\mathcal{N}$ stands for the gaussian/normal probability density function
and the $\mu$ and $\sigma$ represent the mean and standard
deviation. We generate multiple sets of parameters by sampling each gaussian distribution, $\{\alpha^i,\ n^i,\ \theta_r^i,\ \theta_s^i,\ \theta_i^i\}$, where the subscript ``$i$'' is the iteration number. For each set the  value of $K_s$ minimising the objective function was computed by our inversion procedure presented above. We generated enough sets of parameters such as the histogram of $K_s$ values look like a gaussian function with a stabilized standard deviation.  We used this standard deviation as uncertainty on $K_s$.

We did not
consider the uncertainties on radargram picking, because we evaluated it has a very weak
influence comparing to the other uncertainties considered.

Using our analysis, we found in
the case of falling head infiltration that $K_s$ was equal to $0.12\pm0.01\ cm/min$. This narrow range of possible values is in agreement with disk infiltrometer value, and clearly shows the accuracy of our method. 

\section{Constant Head Infiltration Experiment}

\subsection{Numerical Example}
\subsubsection{Forward Modelling}
\begin{figure}[htpb]

\noindent\includegraphics[width=15cm]{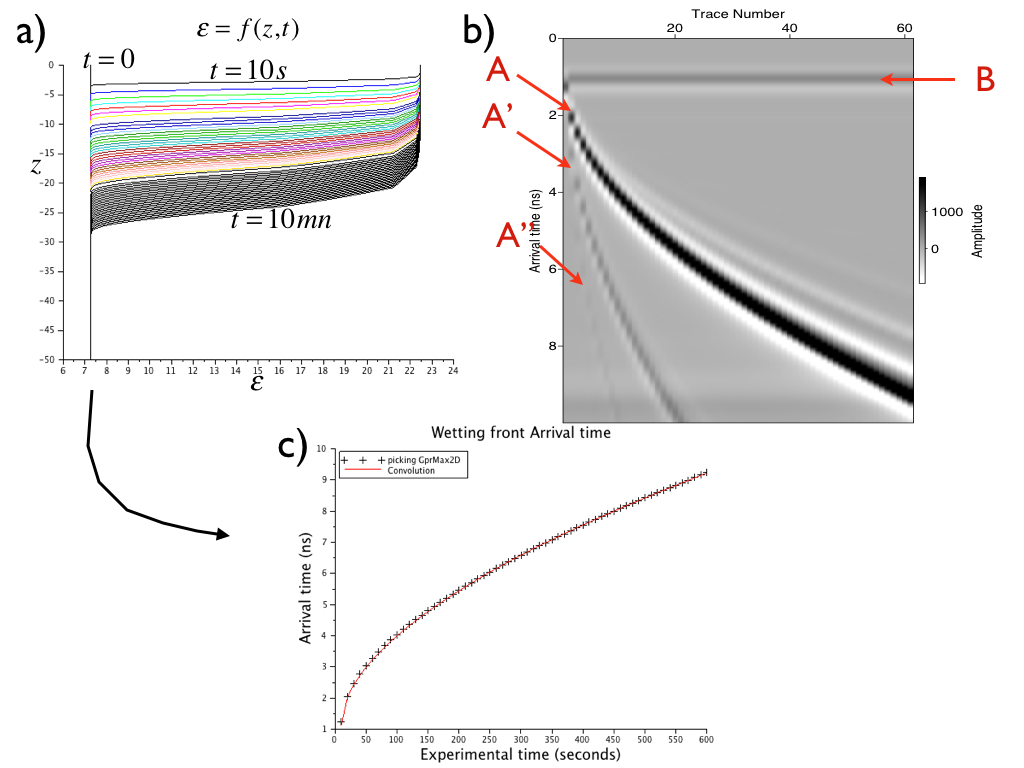}
\caption{Constant head infiltration with 5 cm of water. a)
  Permittivity profiles, each curve is plotted every 10 s. b) Radargram simulated with
  GprMax2D, reflection A is the wetting front, B is the direct wave,
  A' and A'' are multiples. c) Two Way Travel Time computed with our
  convolution algorithm from the simulated permittivity profiles.}
 \label{algo2}
 \end{figure}

In this second case, a water layer of 5 cm above the ground was kept
constant during the entire experiment. Similarly as above, using the
same van Genuchten parameters as in the first synthetic example
(Table~\ref{param}), we modeled infiltration of water inside a ring
infiltrometer by applying a constant pressure head of 5 cm to the top
node during 10 minutes. The permittivity profiles are presented in
Fig.~\ref{algo2}-a, with each curve plotted every 10 s as in the
previous case. Fig.~\ref{algo2}-b shows the radargram simulated
with GprMax2D. As can be seen, the reflection labeled $A$ describing
the position of the infiltration front, is returning
at increasing times, because infiltration is being constantly fed by
the constant ponding depth, contrary to the previous falling head
case. In Fig.~\ref{algo2}-c, we computed the TWT time of the wetting
front using the convolution algorithm and picking the $A$
reflection from the radargram in Fig.~\ref{algo2}-b.
\subsubsection{Inverse Modelling}
We inverted for the
saturated hydraulic conductivity by minimizing the differences
between the arrival times of the wetting front reflection obtained by
the convolution algorithm and the arrival times picked from the simulated radargram
in Fig.~\ref{algo2}-b. The objective function was minimized for
$K_s=0.119\ cm/min$, to be compared with the value used for
simulating the data: $K_s=0.120\ cm/min$.

\subsection{Field Experiment}

\begin{figure}[htpb]

\includegraphics[width=15cm]{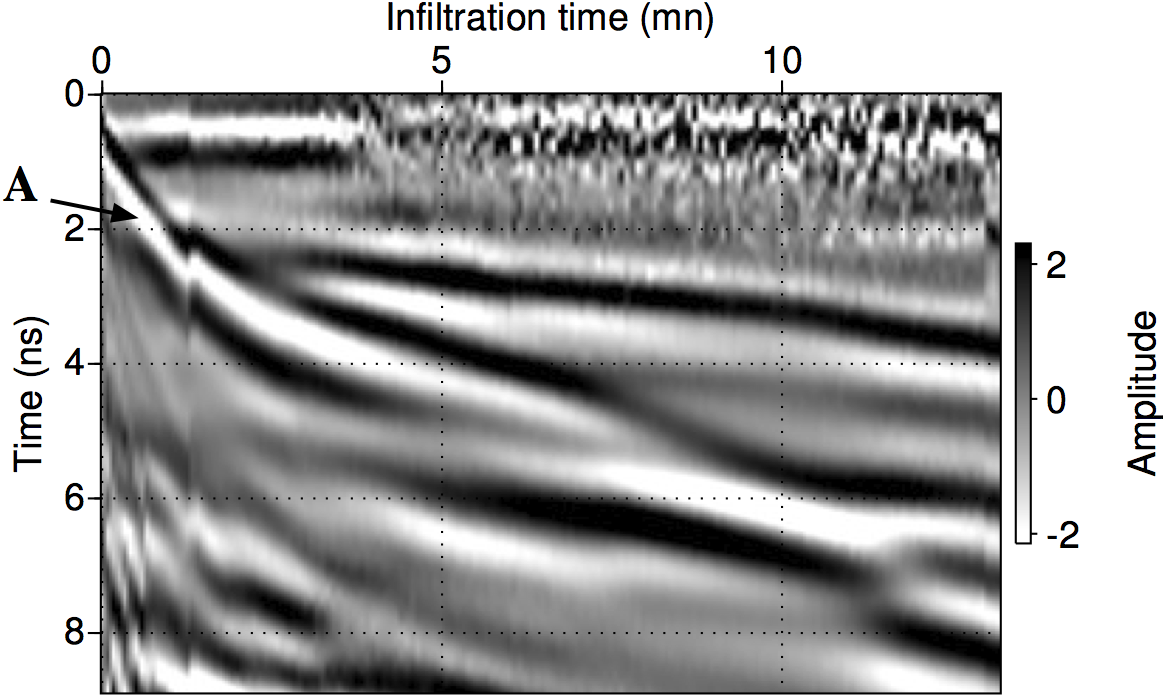}
\caption{GPR data acquired during a constant head (5 cm) infiltration. Reflection A is the reflection coming from the wetting front.}
 \label{R2}
 \end{figure}

The experiment took place in the same quarry of Fontainebleau sand as
the previous experiment. The middle of the antennae was positioned in the
middle of the ring (X = 30 cm in Fig.~\ref{Schema}). The GPR
data are shown in Fig.~\ref{R2} and were recorded during 80 minutes (only a part of the radargram is presented).
We used the van Genuchten parameters determined in the laboratory
using the hanging column experiments (Table~\ref{param}) and we
measured on sand core samples an initial volumetric water content of
$\theta_i=0.07\pm0.02$.

In the profile presented in Fig.~\ref{R2},
the arrival time of 
reflection $A$ ranges from 0 at the beginning of the experiment to
about 6 ns after 10 min. We picked the arrival time of the $A$ reflection peak and computed the objective function using the same procedure
as described before. We obtained the minimum of the objective function
for $K_s=0.089\ cm/min$. Again, this value has to be compared with the
one obtained by the disk infiltrometer experiment,
$K_{Disk}=0.108\pm0.01\ cm/min$. Using the same procedure as
presented in the earlier field example, we found a range of possible
values for the saturated hydraulic conductivity, $K_s =0.089\pm0.005\ cm/min$. Despite the fact that we are not in the same range as the disk infiltrometer method the discrepancy is very small and allows us to conclude on the good accuracy of our method.

\section{Discussion}

The results presented above indicate clearly that a commercial surface GPR can be used as a tool for monitoring the wetting front.
Although the use of surface-based GPR data to estimate the parameters
of unsaturated flow models is not new~\citep{Moysey2010},
our method gives accurate values of the saturated hydraulic
conductivity with uncertainties comparable or smaller than those obtained with disk infiltrometer measurements. A distinct advantage of our approach is the simplicity
of the algorithm and its rapidity to converge, which is very
encouraging for more complicated models ( stack of non-homogeneous layers).

The discrepancy between saturated hydraulic conductivity determined
by disk infiltrometry and that obtained with our GPR algorithm comes from different
phenomena. First of all, the van Genuchten parameters determined from
the water hanging column experiment are obtained with saturation coming from the bottom of the soil samples, whereas in
our case, the infiltration is a ponded one, thus coming from the top.

Despite the fact that we upgraded the single ring infiltrometer by the use of 
GPR to monitor the wetting front, we still suffer from the problem
of entrapped-air, which causes reduction of saturated water content
and hydraulic conductivity. This issue cannot be fixed with
ponded infiltration. Disk infiltrometer measurement
monitoring may cause less problems, working with negative matric
potentials~\citep{Ankeny1991,Reynolds1991}.

During our modeling, we considered our soil as an homogeneous and
isotropic one. Real soils exhibit heterogeneities, triggering preferential flows. Even in the case of our quarry of
Fontainebleau sand, differences in packing and compaction could
lead to creation of preferential flow paths. 

One of the way to solve
this issue could be to use a dual porosity model~\citep{Gerke1993} and a
Monte Carlo procedure to generate a high number of soil models with different 
parameters, as we did with the single porosity model in Hydrus-1D, and performed statistical analysis on the saturated
hydraulic conductivity obtained.

An other source of error, already discussed above, comes from the assumption
that a 3D infiltration monitored by 3D electromagnetic waves can be
treated as a 1-D temporal convolution. This limitation will be studied
in future works, using Hydrus 2D/3D to simulate 2D axisymmetrical
infiltration and 2D infiltration.

The results represent a promising step toward application of multi-parameters inversions. A first study in that direction was presented in
L\'eger~\citep{Leger2013}.

\section{Summary}

This research investigated the use of on-ground surface GPR to monitor the
wetting front during infiltration inside a ring infiltrometer. We
showed by modeling and experiments that a standard GPR device was able
to monitor the displacement of the water front in the soil. We tested in synthetic cases the ability of our algorithm to invert the saturated hydraulic conductivity, knowing the other van Genuchten parameters and the initial water content. Two infiltration experiments were performed, falling head infiltration and constant head infiltration, in a quarry of Fontainebleau sand. The retrieved saturated hydraulic conductivity was comparable to that obtained with disk infiltrometer experiments. Uncertainty analysis accounting for all the van Genuchten parameters, was performed using a modified Monte Carlo method, and proved the robustness of our algorithm. Although results retrieved with GPR were in agreement with disk infiltrometry tests, we stress that further research is needed to improve our algorithm so as to determine the whole set of soil hydrodynamic parameters.

 \bibliographystyle{agu08}

\end{article}
\begin{table}
\begin{tabular}{*{8}{|@{\,}l@{\,}}|}
  \hline
  \multicolumn{8}{|c|}{Falling Head Infiltration}\\
  \hline
  & $\theta_i$ & $\theta_r$ & $\theta_s$& $\alpha$
  & $n$ & $K_s$  & Retrieved $K_s$ \\
  & & & & (cm$^{-1}$) &  & (cm/min) & (cm/min)\\
  \hline
  Numerical & 0.17 & 0.07 & 0.43 & 0.019
  & 8.67 & 0.120 & 0.121 \\
  Field  & 0.09 $\pm$0.01 & 0.062 $\pm$0.003 & 0.39 $\pm$0.01 &
  0.023 $\pm$0.001 & 6.7 $\pm$0.3 & 0.108 $\pm$0.01$^*$ & 0.120 $\pm$0.013 \\
  \hline
  \multicolumn{8}{|c|}{Constant Head Infiltration}\\
  \hline
  & $\theta_i$ & $\theta_r$ & $\theta_s$& $\alpha$
  & $n$ & $K_s$  & Retrieved $K_s$ \\
    & & & & (cm$^{-1}$) &  & (cm/min) & (cm/min)\\
  \hline
  Numerical & 0.17 & 0.07 & 0.43 & 0.019 & 8.67 & 0.120
  &0.119 \\
  Field & 0.07 $\pm$0.02 &  0.062 $\pm$0.003 & 0.39 $\pm$0.01 &
  0.023 $\pm$0.001 & 6.7 $\pm$0.3 & 0.108 $\pm$0.01$^*$ & 0.089
  $\pm$0.009\\
  \hline
\end{tabular}
\caption{Hydrodynamic parameters for the numerical and field
    experiment. The $^*$ indicates values of $K_s$ measured from
    disk infiltrometer experiments.}
\label{param}
\end{table}


\end{document}